\documentclass[aps,showkeys,showpacs,twocolumn,prd]{revtex4}
\usepackage{graphicx}
\usepackage{amsfonts}
\usepackage{amssymb}
\usepackage{epsfig}
\usepackage[anne,small]{caption}
\usepackage{subfig}
\usepackage{bm}
\usepackage{dcolumn}

\def\bea{\begin{eqnarray}}
\def\eea{\end{eqnarray}}
\def\bx{\bm{x}}
\def\bphi{\bm{\phi}}
\def\rmd{\text{d}}

\begin{document}
\title{Hexagonal Structure of Baby Skyrmion Lattices}
\date{\today}

\author{Itay Hen}
\email{itayhe@post.tau.ac.il}
\author{Marek Karliner}
\email{marek@proton.tau.ac.il}
\affiliation{Raymond and Beverly Sackler School of Physics and Astronomy 
Tel-Aviv University, Tel-Aviv 69978, Israel}

\begin{abstract}
We study the zero-temperature crystalline structure of baby Skyrmions
by applying a full-field numerical minimization algorithm 
to baby Skyrmions placed inside different parallelogramic unit-cells 
and imposing periodic boundary conditions.
We find that within this setup, the minimal energy
is obtained for the hexagonal lattice, and that in the resulting configuration 
the Skyrmion splits into quarter-Skyrmions. 
In particular, we find that the energy in the hexagonal case 
is lower than the one obtained on the well-studied rectangular lattice,
in which splitting into half-Skyrmions is observed.
\end{abstract}

\maketitle

\section{Introduction}
The Skyrme model \cite{Skyrme1} is a non-linear
theory for pions in (3+1) dimensions
with topological soliton solutions called Skyrmions. 
The model can be used to describe, with due care \cite{fewNuc1,fewNuc2,fewNuc3}, systems of a few nucleons, and
has also been applied to nuclear and quark matter. One of the most complicated aspects
of the physics of hadrons is the behavior of the phase diagram of hadronic matter at finite
density at low or even zero temperature. 
Particularly, the properties of zero-temperature
Skyrmions on a lattice are of interest, 
since the behavior of nuclear matter at high densities is now a focus of considerable 
interest.
Within the standard zero-temperature Skyrme model description, 
a crystal of nucleons turns
into a crystal of half nucleons at finite density 
\cite{Opt3d,maxSym3D,halfSkyrme0, halfSkyrme1,halfSkyrme2}.
\par
To study Skyrmion crystals one imposes periodic boundary conditions 
on the Skyrme field and works within a unit cell (equivalently, 3-torus) 
$\mathbb{T}_3$ (\cite{TopoSol}, p. 382). The first attempted construction of a crystal was by Klebanov \cite{Opt3d}, 
using a simple cubic lattice of Skyrmions whose symmetries maximize the 
attraction between nearest neighbors. Other symmetries were proposed which 
lead to slightly lower, but not minimal, energy crystals \cite{maxSym3D,halfSkyrme0}.
It is now understood that it is best to arrange the Skyrmions initially as a 
face-centered cubic lattice, with their orientations chosen symmetrically to 
give maximal attraction between all nearest neighbors \cite{halfSkyrme1,halfSkyrme2}. 
\par
The Skyrme model has an analogue in (2+1) dimensions known as the baby Skyrme model,
also admitting stable field configurations of a solitonic nature characterized by integral topological charges
\cite{Old1,Old2}.
Due to its lower dimension, it serves as a simplification of the original model,
but nonetheless it has a physical relevance in its own right,
as a variant of the model arises in ferromagnetic quantum Hall systems \cite{Condensed,QHE,QHF1,QHF2,QHF3}.
This effective theory is obtained when the excitations relative to 
ferromagnetic quantum Hall states are described, in terms of a gradient expansion in the spin density, a field with properties analogous to
the pion field in the $3$D case \cite{Lee}.
\par
The baby Skyrme model has also been
studied in connection with baby Skyrmion lattices under various settings
\cite{2DLattice1,2DLattice2,WC1,WC2,WC3}  
and in fact, it is known that the baby Skyrmions also 
split into half-Skyrmions when placed inside a rectangular lattice \cite{WC1}.
However, to our knowledge, to date it hasn't been known if the rectangular periodic boundary conditions 
yield the true minimal energy configurations over all possible lattice types.
Conversely, if there are other non-rectangular baby Skyrmion lattice configurations
which have lower energy. Finding the answers to these questions is thus 
of particular importance both because of their relevance to
quantum Hall systems in two-dimensions, and also because they may be used to conjecture the
crystalline structure of nucleons in three-dimensions. 
\par
In two dimensions there are five lattice types
as given by the crystallographic restriction theorem \cite{crt},
in all of which the fundamental unit cell is a certain type of a parallelogram.
To find the crystalline structure of the baby Skyrmions,
we place them 
inside different parallelograms with periodic boundary conditions
and find the minimal energy configurations over all parallelograms of
fixed area (thus keeping the Skyrmion density fixed).
As we show later, there is a certain type of parallelogram, namely the hexagonal,
which yields the minimal energy configuration.
In particular, its energy is lower than the known `square-cell' configurations in which the
Skyrmion splits into half-Skyrmions. As will be pointed out later,
the hexagonal structure revealed here is not unique to the
present model, but also arises in
other solitonic models, such as Ginzburg-Landau vortices \cite{GL},
quantum Hall systems \cite{QHE,QHF1}, and even in the context of $3$D Skyrmions \cite{3Dhexlat}.
The reason for this will be discussed in the concluding section. 
\par
In the following section we review the setup of
our numerical computations, introducing
a systematic way for the detection of the minimal energy crystalline structure
of baby Skyrmions. In subsequent sections
we outline the numerical procedure through which
the full field minimal configurations are obtained.
In section \ref{sec:res} we present the main results of our study
and in section \ref{sec:semiAna}, a somewhat more analytical 
analysis of the problem is presented. 
In the last section we make some remarks with regards to future research.

\section{Baby Skyrmions inside a parallelogram}
\begin{figure}[h!]
\includegraphics[angle=0,scale=1,width=0.47\textwidth]{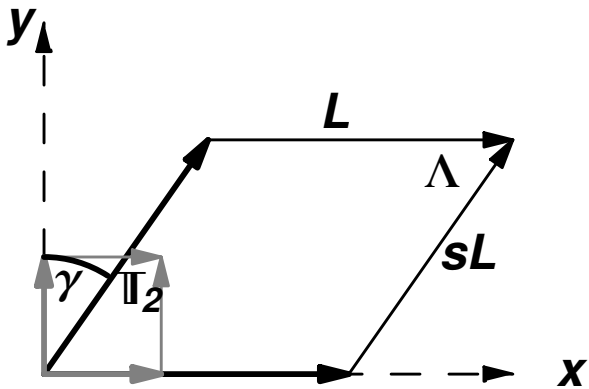}
\caption{\label{par}The parameterization of the fundamental unit-cell parallelogram $\Lambda$ (in black)
and the two-torus $\mathbb{T}_2$ into which it is mapped (in gray).$\hfill{}$}

\end{figure}

The target manifold in the baby Skyrme model is described by a three-dimensional vector 
$\bphi$  with the constraint $\bphi \cdot \bphi=1$ (\textit{i.e.}, $\bphi \in S_2$)
and its Lagrangian density is:
\bea \label{eq:BabyLag}
  \mathcal{L}&=&\frac1{2} \partial_{\mu} \bphi \cdot \partial^{\mu} \bphi 
 - \frac{\kappa^2}{2}\big[(\partial_{\mu} \bphi \cdot \partial^{\mu} \bphi)^2 \\\nonumber
 &-& (\partial_{\mu} \bphi \cdot \partial_{\nu} \bphi) 
\cdot (\partial^{\mu} \bphi \cdot \partial^{\nu} \bphi)
\big] - \mu^2 (1 -\phi_3)
\eea
consisting of a kinetic term, a $2$D Skyrme term,
and a potential term.
The static solutions of the model are
those field configurations which minimize the static energy
functional:
\bea
 E =\frac1{2} \int_{\Lambda}  \rmd x \rmd y &
\Big((\partial_{x} \bphi)^2+ (\partial_{y} \bphi)^2 &
+  \kappa^2 (\partial_x \bphi \times \partial_y \bphi)^2
\nonumber \\ &+ 2 \mu^2 (1 -\phi_3) 
\Big)&
\eea
within each topological sector, where, in our setup, the integration is over parallelograms denoted here
by $\Lambda$:  
\bea
\Lambda = \{\alpha_1 (L,0) + \alpha_2 (s L \sin \gamma ,s L \cos \gamma): 0 \leq \alpha_1,\alpha_2<1 \}\,.
\eea
Here $L$ is the length of one side of the parallelogram, $s L$ with $0 < s \leq 1$ is the length of its other side 
and $0 \leq \gamma < \pi/2$ is the angle between  the `$s L$' side and
the vertical to the `$L$' side (as sketched in Fig. \ref{par}). 
Each parallelogram is thus specified by a set $\{L,s,\gamma \}$ and 
the Skyrmion density inside a parallelogram is  $\rho = B/(s L^2 \cos \gamma)$,
where $B$ is the topological charge of the Skyrmion. 
The periodic boundary conditions are taken into account by 
identifying each of the two opposite sides of a parallelogram as equivalent:
\bea
\bphi(\bx)=\bphi(\bx + n_1 (L,0) +n_2 (s L \sin \gamma ,s L \cos \gamma)) \,,
\eea
with $(n_1,n_2 \in \mathbb{Z})$. We are interested in static finite-energy solutions, which in the language of differential
geometry are $\mathbb{T}_2 \mapsto S_2$ maps. These 
are partitioned into homotopy sectors parameterized by
an invariant integral topological charge $B$, the degree of the map, given by:
\bea \label{eq:O3Bnumber}
 B=\frac1{4 \pi} \int_{\Lambda} \rmd x \rmd y 
\left( \bphi \cdot ( \partial_{x} \bphi \times  \partial_{y} \bphi)
\right) \,.
\eea
The static energy $E$ can be shown to satisfy 
\bea \label{ineq}
E \geq 4 \pi B
\eea
with equality possible only in the `pure' $O(3)$ case (\textit{i.e.}, when both 
the Skyrme and the potential terms are absent) \cite{WC1}. 
We note that while in the baby Skyrme model on $\mathbb{R}^2$ with fixed boundary conditions
the potential term is necessary to prevent the solitons from expanding indefinitely,
in our setup it is not required, due to
the periodic boundary conditions \cite{WC1}.
In this work we study the model both with and without the potential term.
\par
The problem in question can be simplified by a linear mapping of
the parallelograms $\Lambda$ into the 
unit-area two-torus $\mathbb{T}_2$ (see Fig. \ref{par}).
In the new coordinates, the energy functional becomes
\begin{widetext}
\bea \label{eq:T2energy}
E=
\frac{1}{2 s \cos \gamma} 
\int_{\mathbb{T}_2} \rmd x \rmd y 
\left( s^2 (\partial_{x} \bphi)^2
-2 s \sin \gamma  \partial_{x} \bphi \partial_{y} \bphi
+ (\partial_{y} \bphi)^2 \right) +
\frac{\kappa^2 \rho}{2 B} \int_{\mathbb{T}_2} \rmd x \rmd y 
\left( \partial_x \bphi \times \partial_y \bphi \right)^2
+ \frac{\mu^2 B}{\rho}  \int_{\mathbb{T}_2} \rmd x \rmd y \left(1 -\phi_3 \right)
\,.
\eea
\end{widetext}
Note that the dependence of the energy on the Skyrme parameters $\kappa$ and $\mu$ and the Skyrmion density 
$\rho$ is only through $\kappa^2 \rho$ and $\mu^2/ \rho$. 
\par
In order to find the minimal energy configuration of Skyrmions over all
parallelograms with fixed area (equivalently, a specified $\rho$),
we scan the parallelogram parameter space  $\{s,\gamma \}$ and find the parallelogram 
for which the resultant energy is minimal over the parameter space. 
An alternative approach to this problem, which is of a more analytical nature,
has also been implemented, and is 
discussed in detail in section \ref{sec:semiAna}. 
In the following section we describe the numerical minimization procedure.

\section{The numerical minimization procedure}
The Euler-Lagrange equations
derived from the energy functional (\ref{eq:T2energy})
are non-linear \textit{PDE}'s, so in general one must resort to 
numerical techniques. The minimal energy configuration of a Skyrmion with charge $B$
and a given parallelogram (defined by a set $\{ s,\gamma \}$) with a
specified Skyrmion density $\rho$
is found by a full-field relaxation method on 
a $100 \times 100$ rectangular grid on the two-torus $\mathbb{T}_2$,
where at each point a field triplet $\bphi(x_m,y_n)$ is defined. 
For the calculation of the energy and charge densities, we
use the procedure devised by \cite{Hale},
in which the evaluation of these quantities is performed at the centers 
of the grid squares.
Numerical derivatives are also evaluated at these points; 
the $x$-derivatives are  calculated by  
\bea
 &&\frac{\partial\bphi}{\partial x}\Big|_{(x_{m+\frac{1}{2}},y_{n+\frac{1}{2}})} =  \\\nonumber
\frac1{\Delta x}  & \Bigg(  &
  {\Big \langle} \frac{\bphi(x_{m+1},y_{n}) + \bphi(x_{m+1},y_{n+1})}{2}  {\Big \rangle}_{\rm normed} \\\nonumber
 & - &{\Big \langle}\frac{\bphi(x_{m},y_{n}) + \bphi(x_{m},y_{n+1})}{2}  {\Big \rangle}_{\rm normed} 
 \Bigg)\,,
\eea
with the $y$-derivatives analogously defined, and the ``normed'' subscript indicates that the averaged fields
are normalized to one. 
We find this procedure to work very well in practice. For a more detailed discussion
of the method, see \cite{Hale}. 
\par
The relaxation process begins by initializing the field triplet $\bphi$
to a rotationally-symmetric configuration  
\bea \label{eq:initConfig}
\bphi_{\textrm{initial}}=(\sin f(r) \cos B \theta,\sin f(r) \sin B \theta,\cos f(r)) \quad,
\eea
where the profile function $f(r)$ is set to $f(r)=\pi \exp(-r)$
with $r$ and $\theta$ being the usual polar coordinates.
The energy of the baby Skyrmion is then minimized by repeating the following steps:
a point $(x_m,y_n)$ on the grid is chosen at random, along 
with one of the three components of the field $\bphi(x_m,y_n)$.
The chosen component is then shifted by a value $\delta_{\phi}$ chosen uniformly from the segment $[-\Delta_{\phi},\Delta_{\phi}]$
where $\Delta_{\phi}=0.1$ initially. The field triplet is then normalized
and the change in energy is calculated.
If the energy decreases, the modification of the field is accepted
and otherwise it is discarded.
The procedure is repeated, while the value of 
$\Delta_{\phi}$ is gradually decreased throughout
the procedure. This is done until no further decrease in energy is observed.\par
\par 
To check the stability and reliability of our numerical prescription,
we set up the minimization scheme using
different initial configurations and an $80 \times 80$ grid
for several $\rho$, $s$ and $\gamma$ values. 
This was done in order to make sure that the final configurations are independent of the
discretization and of the cooling scheme.
As a further check, some of the minimizations were repeated 
with numerical derivatives of a higher precision, using eight field points for
their evaluation. No apparent changes in the results were detected. 

\section{\label{sec:res} Results}
Using the minimization procedure presented in the previous section, 
we have found the minimal energy static Skyrmion configurations
over all parallelograms, for various settings:
The `pure' $O(3)$ case, in which both $\kappa$, the Skyrme parameter, and $\mu$, 
the potential coupling, are set to zero, the Skyrme case for which only $\mu=0$, and
the general case for which neither the Skyrme term nor the potential term vanish. 
\par
In each of these settings, the parameter space of parallelograms was scanned, 
while the Skyrmion density $\rho$ is held fixed, 
yielding for each set of $\{ s,\gamma \}$ a minimal energy configuration.
The choice as to how many Skyrmions are to be placed inside the unit cells
was made after some preliminary testing in which Skyrmions of other charges (up to $B=8$) were also 
examined. The odd-charge minimal-energy configurations turned out to have 
substantially higher energies than even-charge ones,
where among the latter, the charge-two Skyrmion was found to be the most fundamental,
as it was observed that it serves as a `building-block' for the higher-charge ones.
In what follows, a summary of the results is presented.

\subsection{The pure $O(3)$ case ($\kappa=\mu=0$)}
\begin{figure}[htp!]
\begin{tabular}[b]{|c|c|}
\hline
\subfloat[$s=1$ and $\gamma=0$]{
\includegraphics[angle=0,scale=1,width=0.16\textwidth]{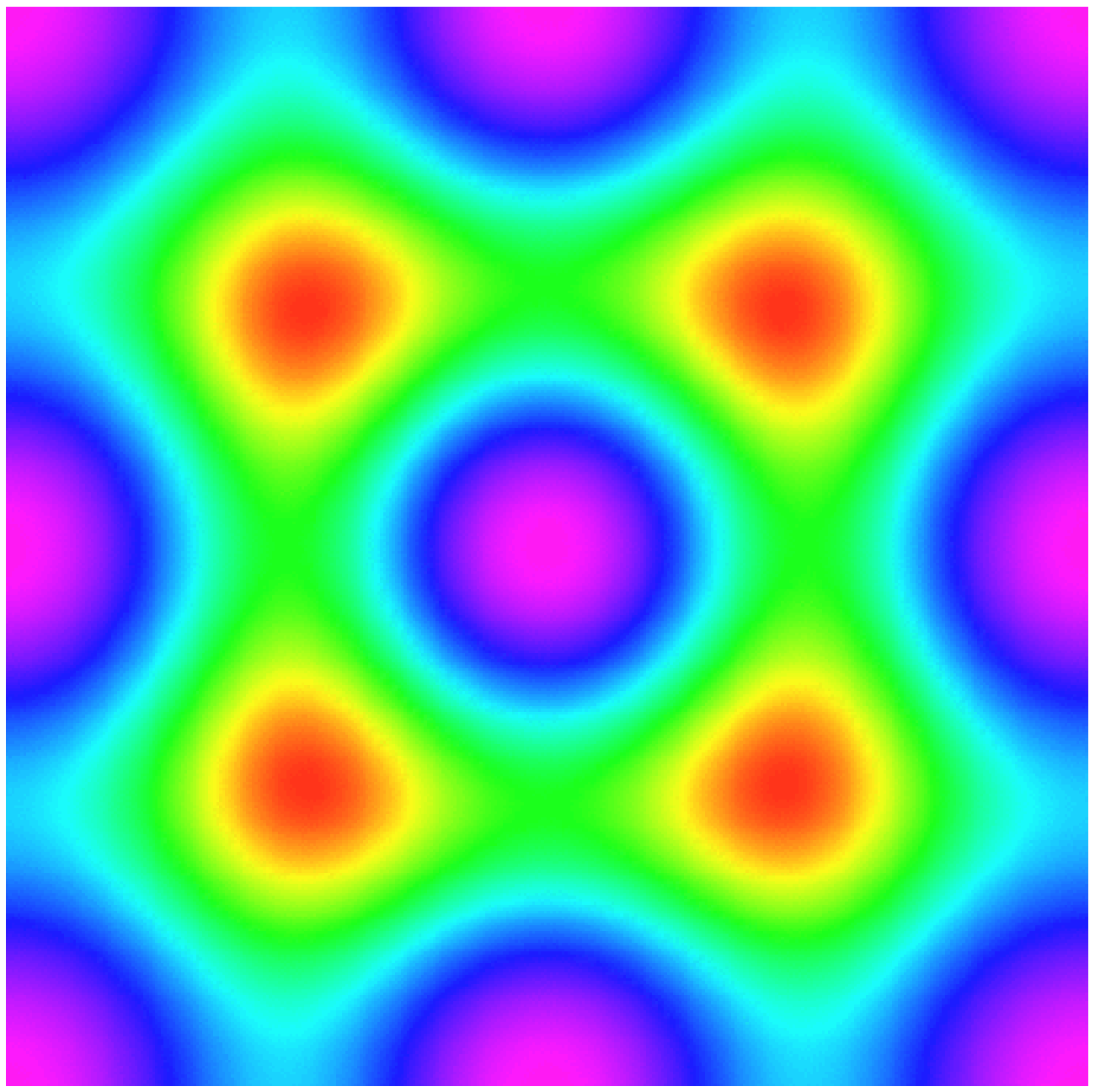}}
&
\subfloat[$s=0.9$ and $\gamma=\pi/16$]{
\includegraphics[angle=0,scale=1,width=0.21\textwidth]{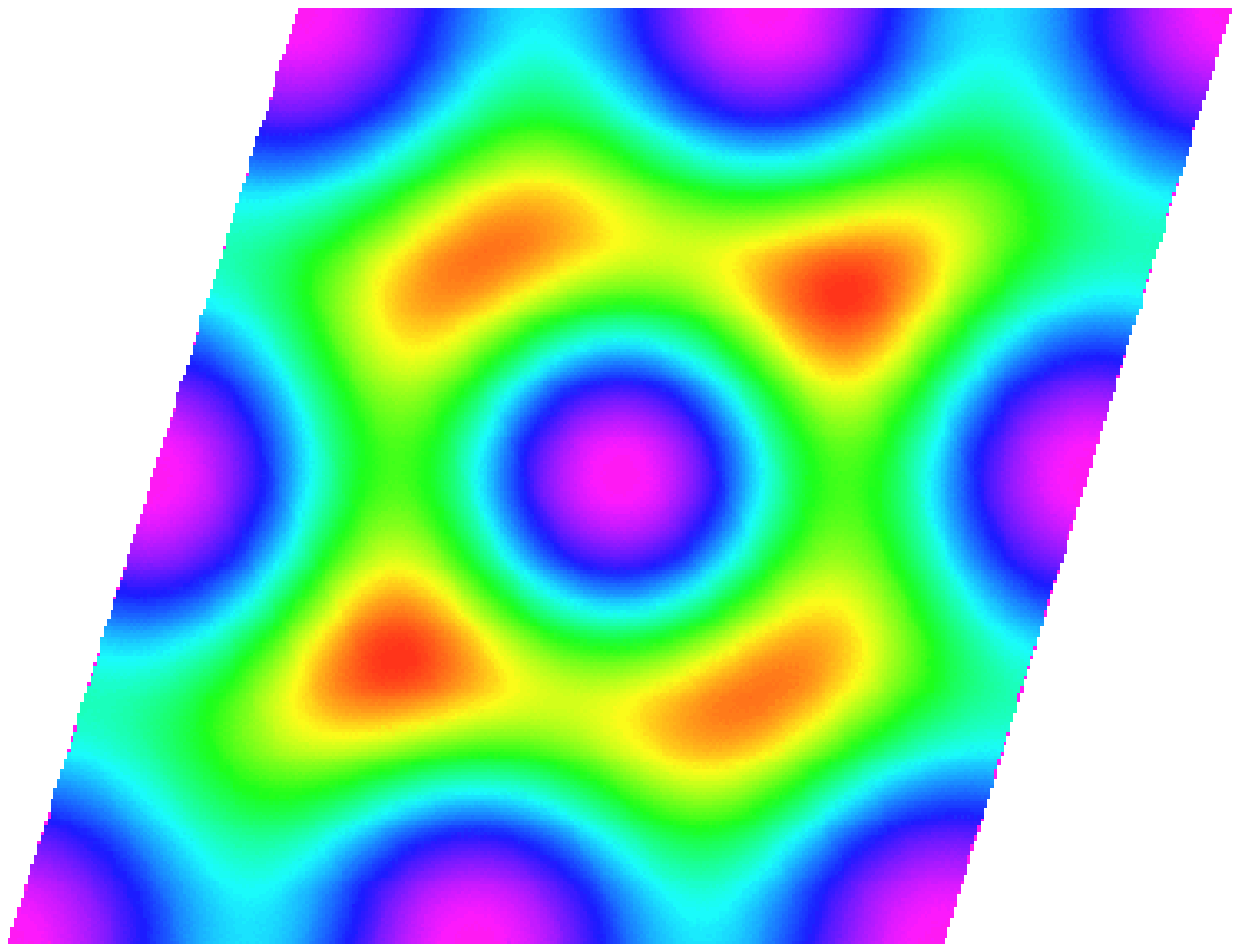}}
\\
\hline
\subfloat[$s=0.7$ and $\gamma=0$]{
\includegraphics[angle=0,scale=1,width=0.185\textwidth]{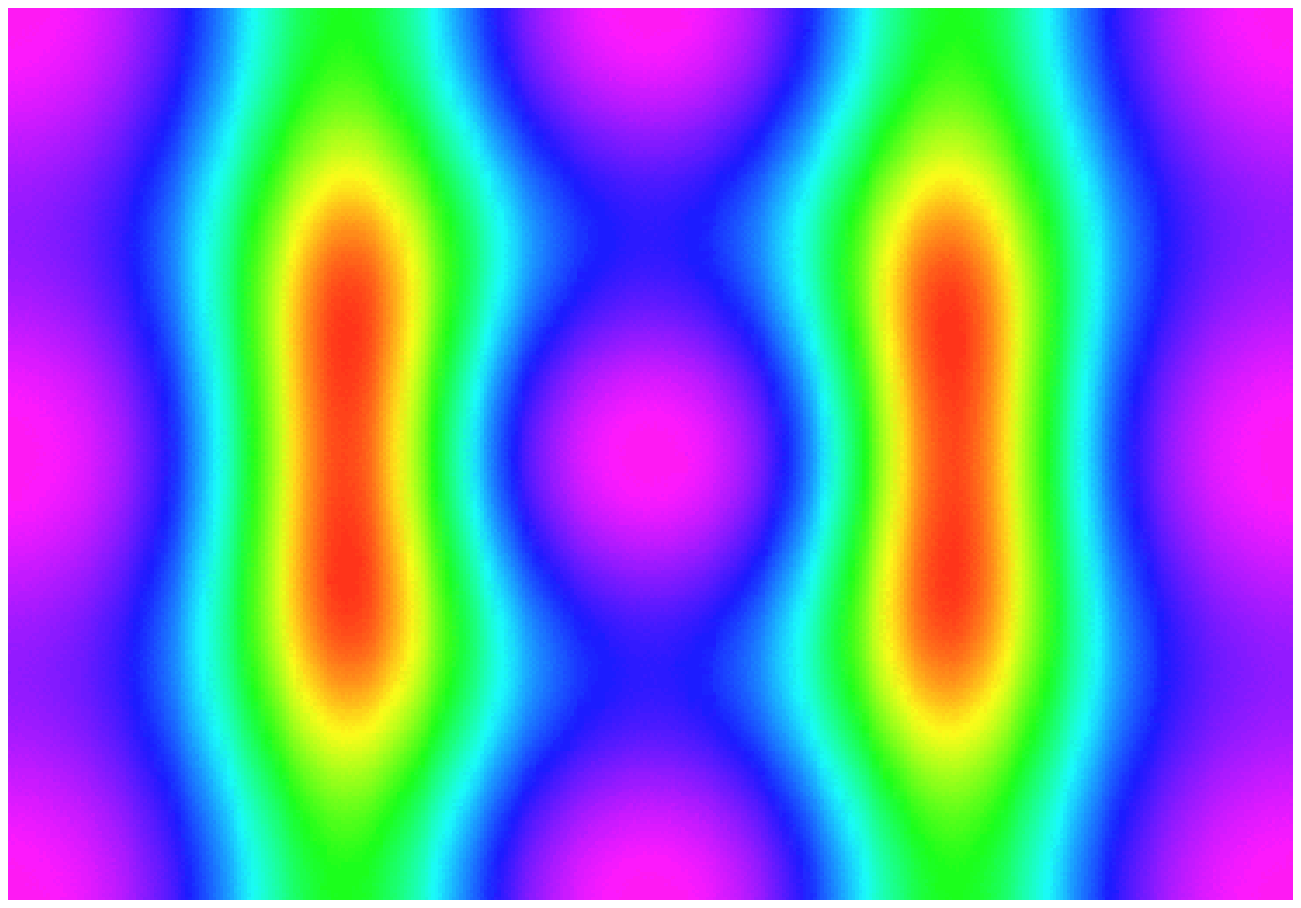}}
&
\subfloat[$s=1$ and $\gamma=\pi/6$]{
\includegraphics[angle=0,scale=1,width=0.235\textwidth]{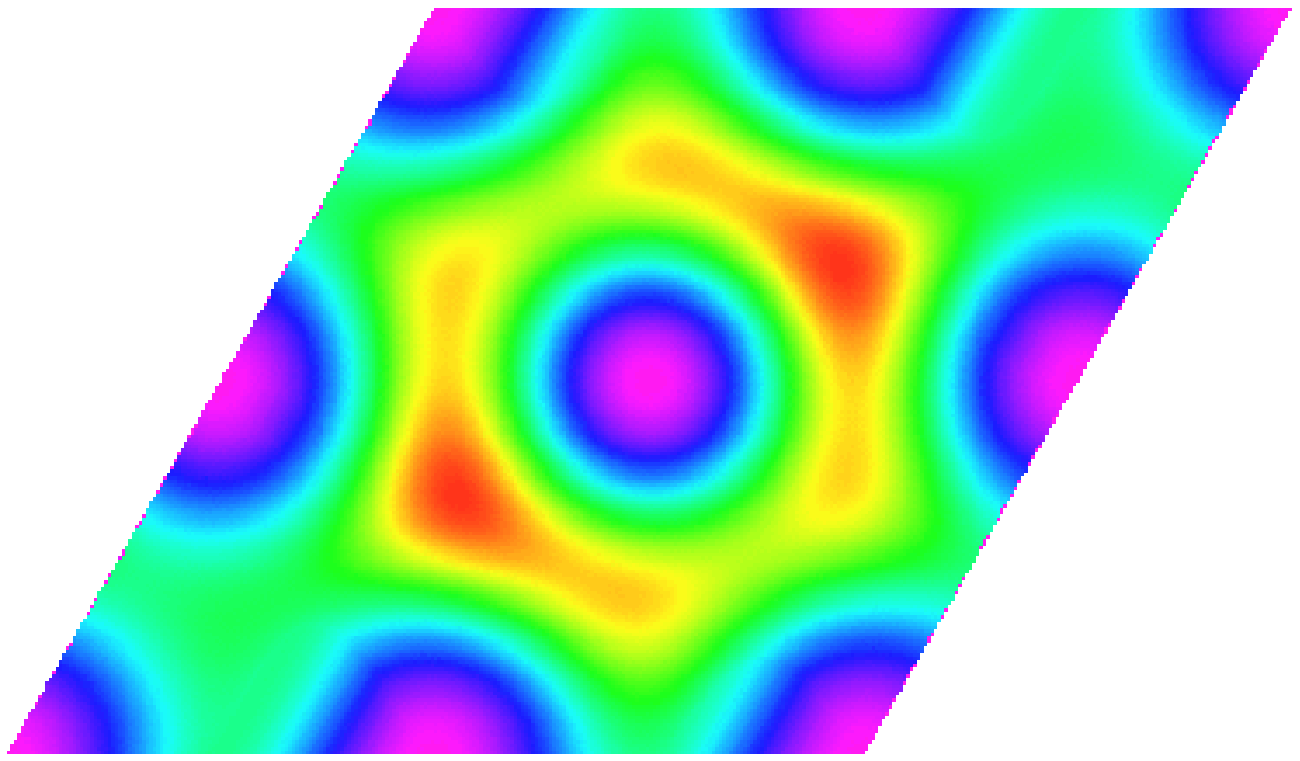}}
\\
\hline
\end{tabular}
\begin{flushleft}
\caption{\label{cK0}Charge-two Skyrmions in the pure $O(3)$ case: 
Contour plots of the charge densities ranging from violet
(low density) to red (high density)
for various parallelogram settings, 
all of which saturate the energy bound $E=4 \pi B=8 \pi$.}
\end{flushleft}\end{figure}

The pure $O(3)$ case corresponds to setting both $\kappa$ and $\mu$
to zero. In this case, analytic solutions in terms of Weierstrass elliptic functions may be 
found \cite{WC1,WC2,WC3} and the minimal energy configurations, 
in all parallelogram settings, saturate the energy bound in (\ref{ineq}) 
giving $E=4 \pi B$. 
Thus, comparison of our numerical results with the analytic solutions serves as a useful check 
on the precision of our numerical procedure.
The agreement is to $6$ significant digits. 
Contour plots of the charge densities for
different parallelogram settings for the charge-two Skyrmions are shown in Fig. \ref{cK0},
all of them of equal energy $E/8 \pi =1$.

\subsection{The Skyrme case ($\kappa \neq 0, \mu = 0$)}
\begin{figure}[hbp!]
\begin{tabular}[b]{|c|c|}
\hline
\subfloat[$s=1$, $\gamma=0$, and \hbox{$E/8 \pi =1.446$}]{
\label{figureB2a} %% label for first subfigure
\includegraphics[angle=0,scale=1,width=0.14\textwidth]{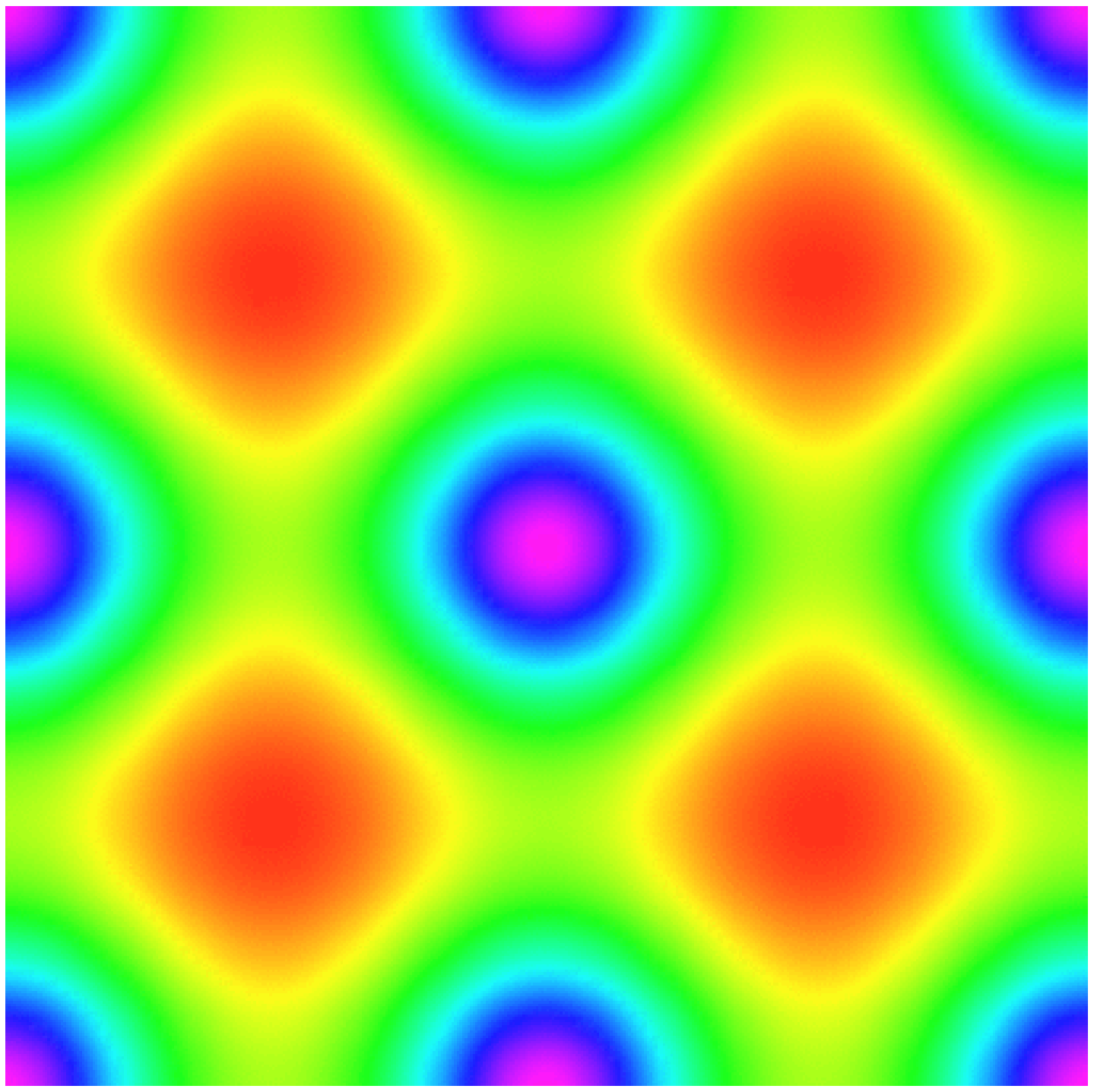}}
&
\subfloat[$s=1$, $\gamma=\pi/6$, and \hbox{$E/8 \pi =1.433$}]{
\label{figureB2b} %% label for first subfigure
\includegraphics[angle=0,scale=1,width=0.23\textwidth]{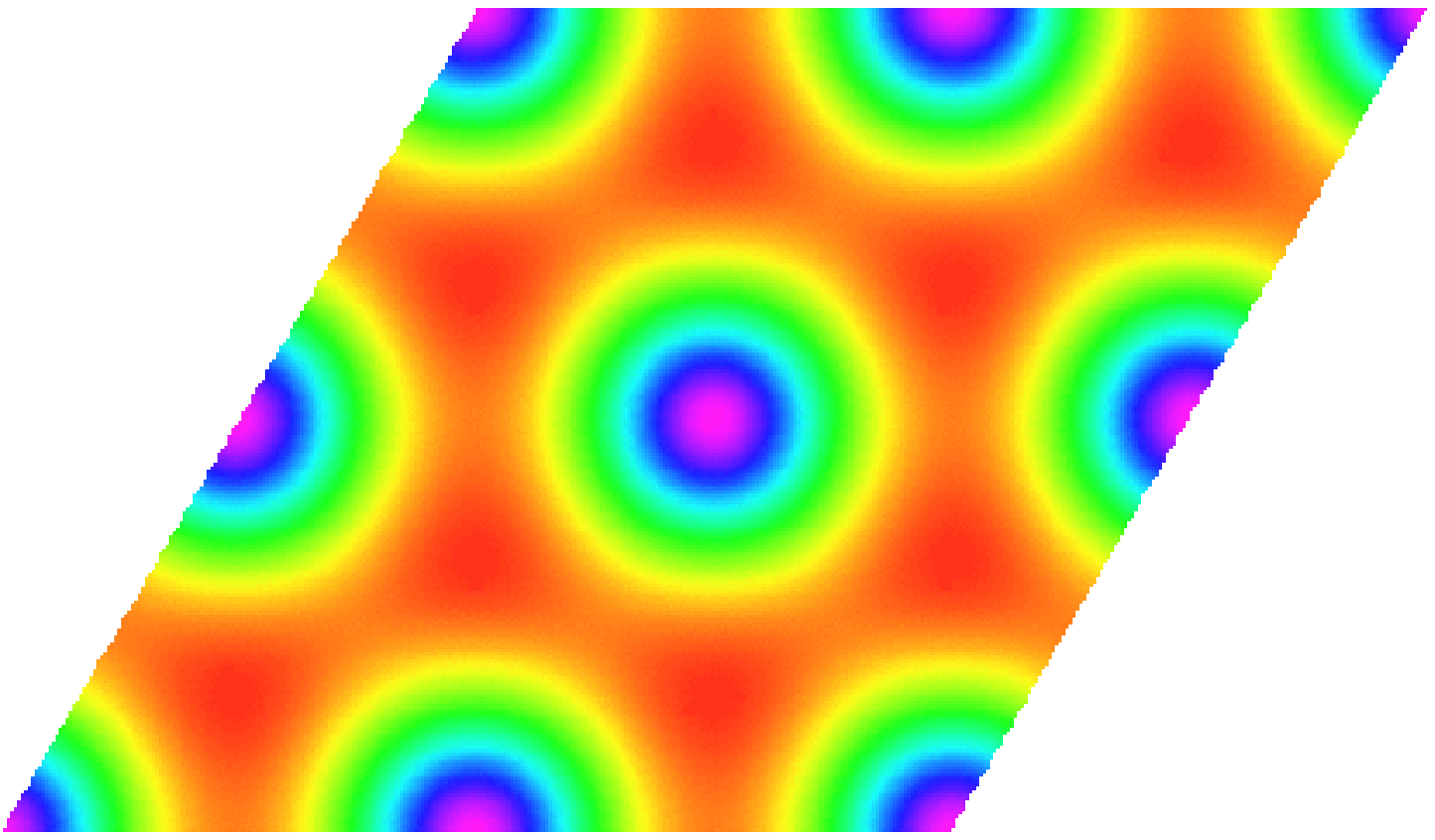}}
\\
\hline
\subfloat[$s=0.51$, $\gamma=0$, and \hbox{$E/8 \pi =1.587$}]{
\label{figureB2c}
\includegraphics[angle=0,scale=1,width=0.2\textwidth]{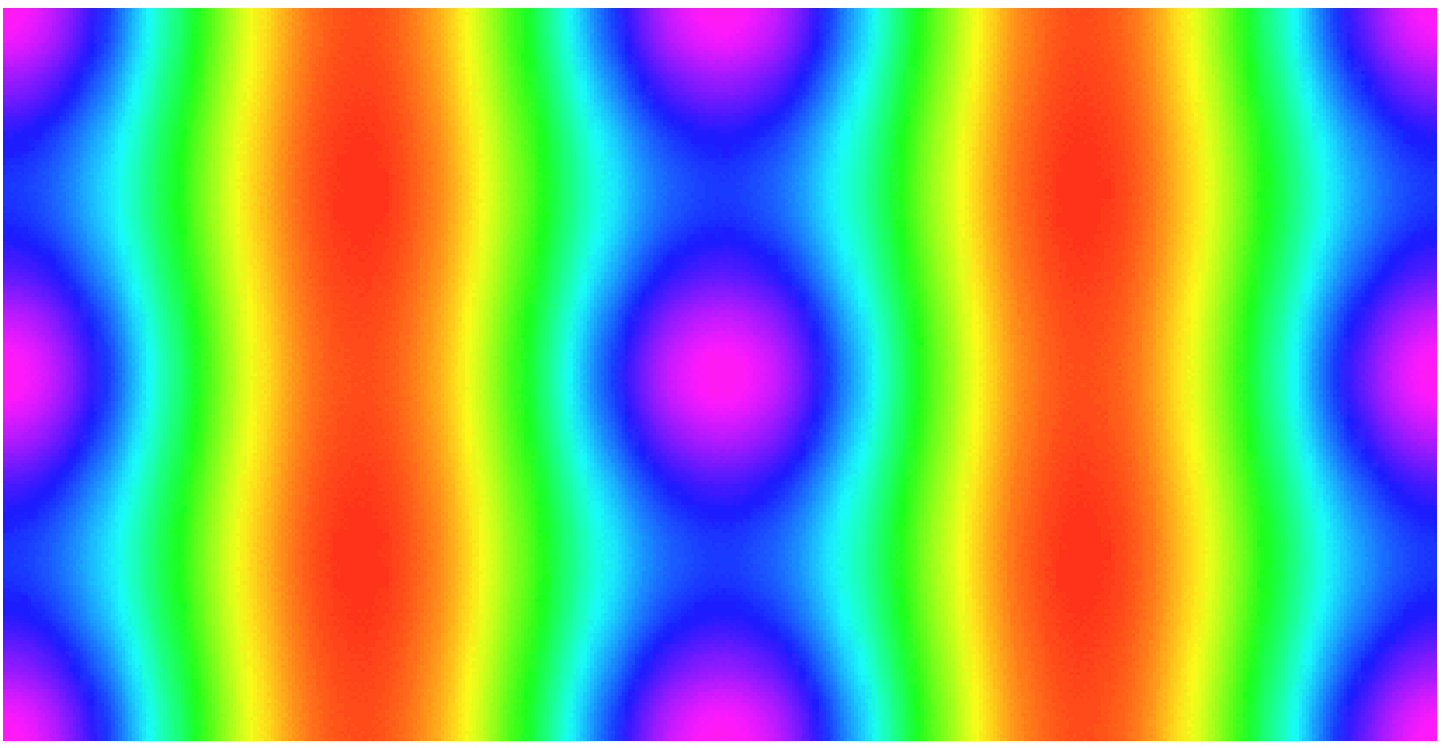}}
&
\subfloat[$s=1$, $\gamma=\pi/4$, and \hbox{$E/8 \pi =1.454$}]{
\label{figureB2d}
\includegraphics[angle=0,scale=1,width=0.24\textwidth]{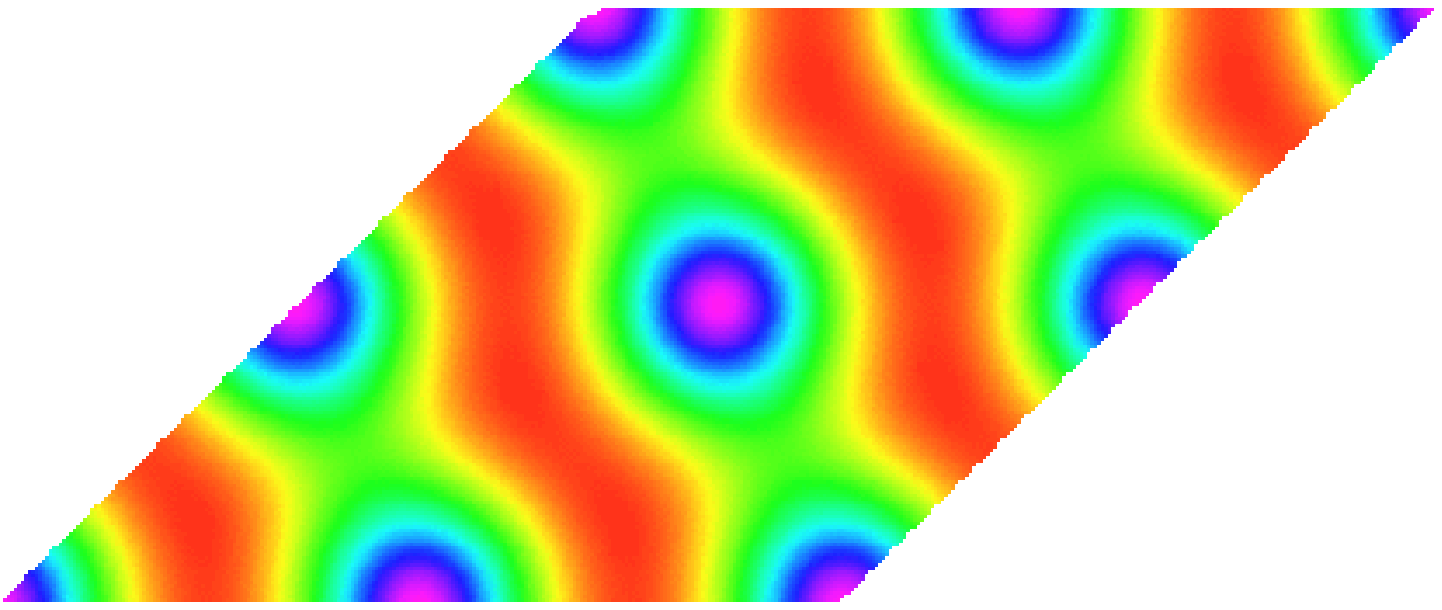}}
\\
\hline
\end{tabular}
\caption{\label{contourB2}Charge-two Skyrmions in the Skyrme case with $\kappa^2=0.03$ and $\rho=2$: 
Contour plots of the charge densities
for the hexagonal, square and other fundamental cells 
ranging from violet (low density) to red (high density).
As the captions of the individual subfigures indicate, the hexagonal configuration  
has the lowest energy.$\hfill{}$}
\end{figure}

As pointed out earlier, in setting $\mu=0$, the dependence of the energy functional on
the Skyrme parameter $\kappa$ is only through $\kappa^2 \rho$, 
so without loss of generality we vary $\rho$ and fix $\kappa^2=0.03$ throughout (this
particular choice for $\kappa$ was made for numerical convenience).
Minimization of the
energy functional (\ref{eq:T2energy}) over all parallelograms yielded the following.
For any fixed density $\rho$, the minimal energy 
was obtained for $s = 1$ and $\gamma =\pi/6$. This set of values 
corresponds to the `hexagonal' or `equilateral triangular' lattice. 
In this configuration, any three adjacent zero-energy loci (or `holes')
are the vertices of equilateral triangles, and 
eight distinct high-density peaks are observed (as shown in Fig. \ref{figureB2b}).
This configuration can thus be interpreted as the splitting of the two-Skyrmion into 
eight quarter-Skyrmions.
This result turned out to be independent of the Skyrmion density $\rho$. 
In particular, the well-studied square-cell minimal energy configuration (Fig. \ref{figureB2a}), in which
the two-Skyrmion splits into four half-Skyrmions
turned out to have higher energy than the hexagonal case. Figure \ref{contourB2}
shows the total energies (divided by $8 \pi$) and 
corresponding contour plots of charge densities of the hexagonal, square and other 
configurations (for comparison), all of them with $\rho=2$.

\begin{figure}[htp!]
\subfloat[]{
\label{energyB} 
\includegraphics[angle=0,scale=1,width=0.44\textwidth]{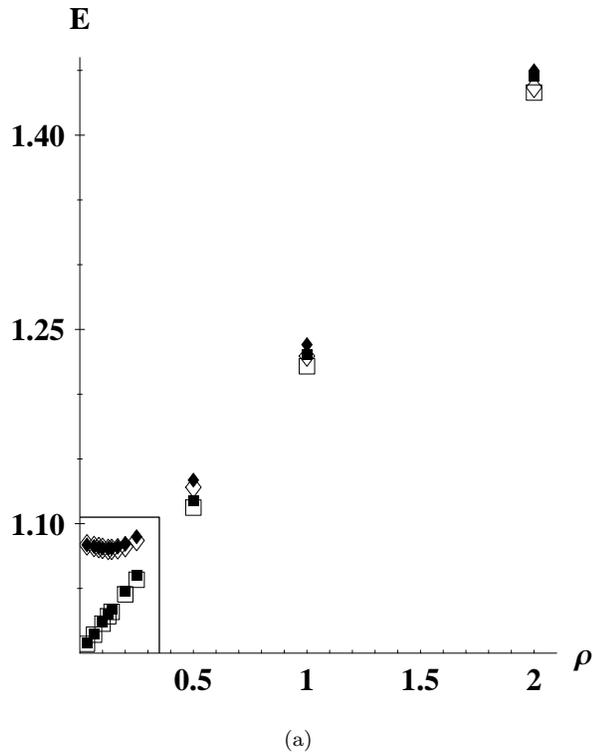}}
\hspace{1cm}
\subfloat[]{
\label{energyA} 
\includegraphics[angle=0,scale=1,width=0.44\textwidth]{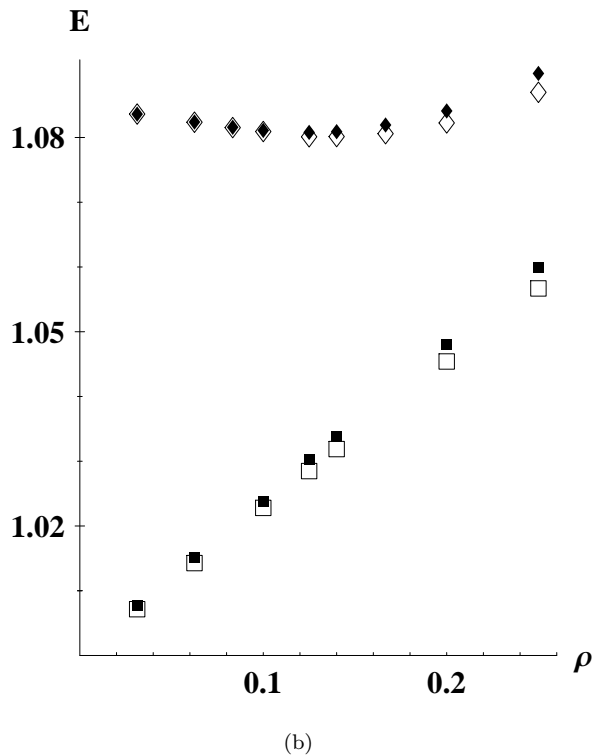}}
\caption{\label{Energy} Total energy $E$ (divided by $8 \pi$) of the charge-two Skyrmion
in the hexagonal lattice ({\small $\square$} -- Skyrme case and {\small $\lozenge$} -- general case)
and in the square lattice ( {\scriptsize $\blacksquare$} -- Skyrme case and {\scriptsize $\blacklozenge$} -- general case) 
as function of the Skyrmion density (in the Skyrme case, $\kappa^2=0.03$ and in the general case 
$\kappa^2=0.03$ and $\mu^2=0.1$).
Note the existence of 
an optimal density (at $\rho \approx 0.14$) in the general case, for which
the energy attains a global minimum.
Figure (b) is an enlargement of the lower left corner of figure (a).
$\hfill{}$}
\end{figure}

The total energy of the Skyrmions in the hexagonal setting  
turned out to be linearly proportional to the density of the 
Skyrmions, reflecting the scale invariance of the model (Fig. \ref{Energy}). 
In particular, the global minimum of $E=4 \pi B =8 \pi$ is reached when $\rho \to 0$.  
This is expected since setting $\rho=0$
is equivalent to setting the Skyrme parameter $\kappa$ to zero, in which case
the model is effectively pure $O(3)$ and inequality (\ref{ineq}) is saturated.
\begin{figure}[hbp!]
\includegraphics[angle=0,scale=1,width=0.5\textwidth]{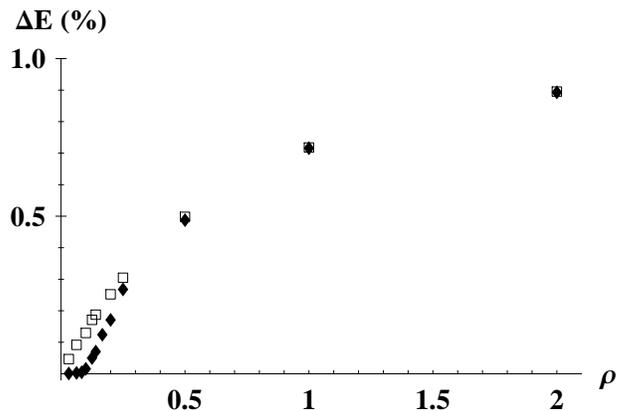}
\caption{\label{EnergyDelta}
Energy difference (in $\%$) between the hexagonal setting and the square setting,
in the Skyrme case ({\scriptsize $\square$}) and in the general case ({\scriptsize $\blacklozenge$}),
as function of the Skyrmion density. As the density decreases, the energy difference 
between the two lattice types is reduced.$\hfill{}$}
\end{figure}
\subsection{\label{subsec:general} The general case ($\kappa \neq 0$, $\mu \neq 0$)}
The hexagonal setting turned out to be the 
energetically favorable in the general case as well.
Since, however, in this case, the Skyrmion possesses a 
definite size (as can be verified by looking at the $\rho$ dependence in the energy functional),
the Skyrmion structure is different at low and at high densities and a phase transition is observed.
While at low densities the Skyrmions stand isolated from one another,
at high densities they fuse together forming the quarter-Skyrmion crystal as in
the Skyrme case reported above.
As the density $\rho$ decreases or equivalently the value of $\mu$ increases,
the size of the Skyrmions becomes small compared to the cell size. 
The exact shape of the lattice loses its effects and the differences in
energy among the various lattice types become very small. This is illustrated in
figures \ref{Energy} and \ref{EnergyDelta}. 
\par
\begin{figure}
\begin{tabular}[b]{|c|c|}
\hline
\subfloat[$\rho=1$, $E/8 \pi =1.229$]{
\label{figureG2a} %% label for first subfigure
\includegraphics[angle=0,scale=1,width=0.22\textwidth]{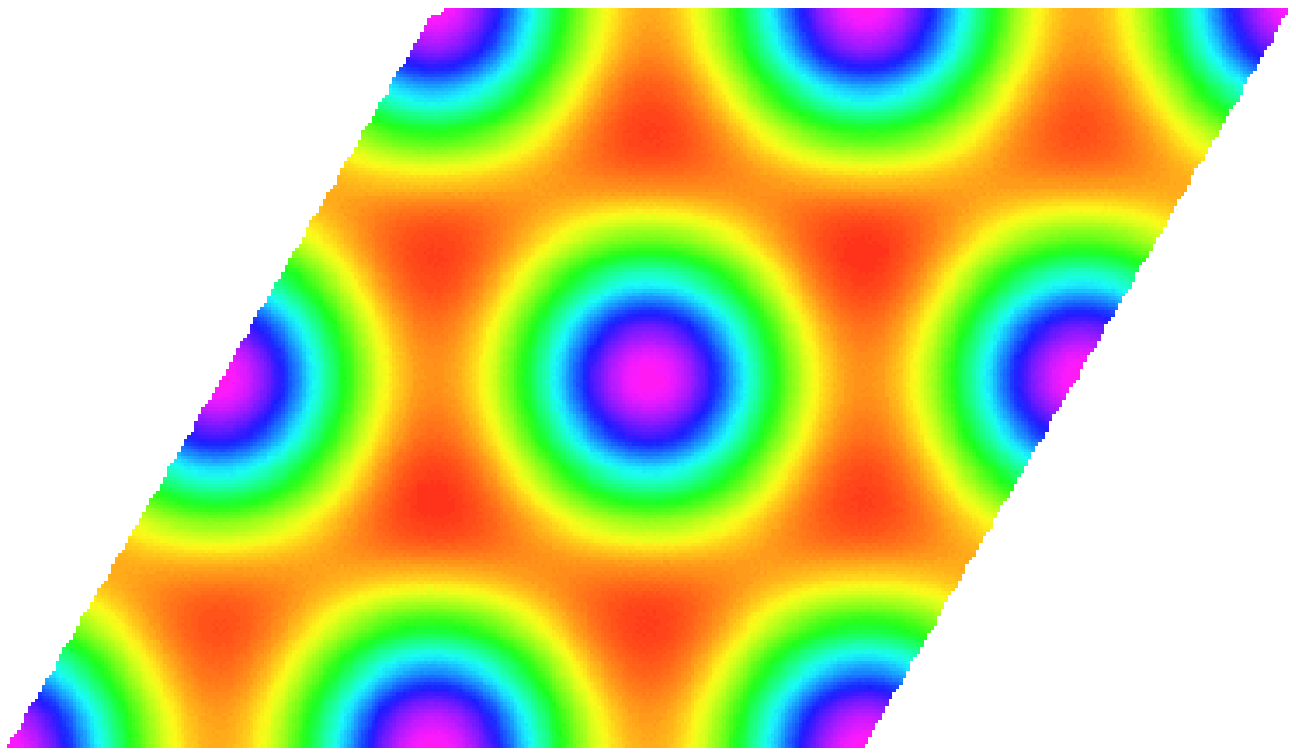}}
&
\subfloat[$\rho=0.5$, $E/8 \pi =1.128$]{
\label{figureG2b} %% label for first subfigure
\includegraphics[angle=0,scale=1,width=0.22\textwidth]{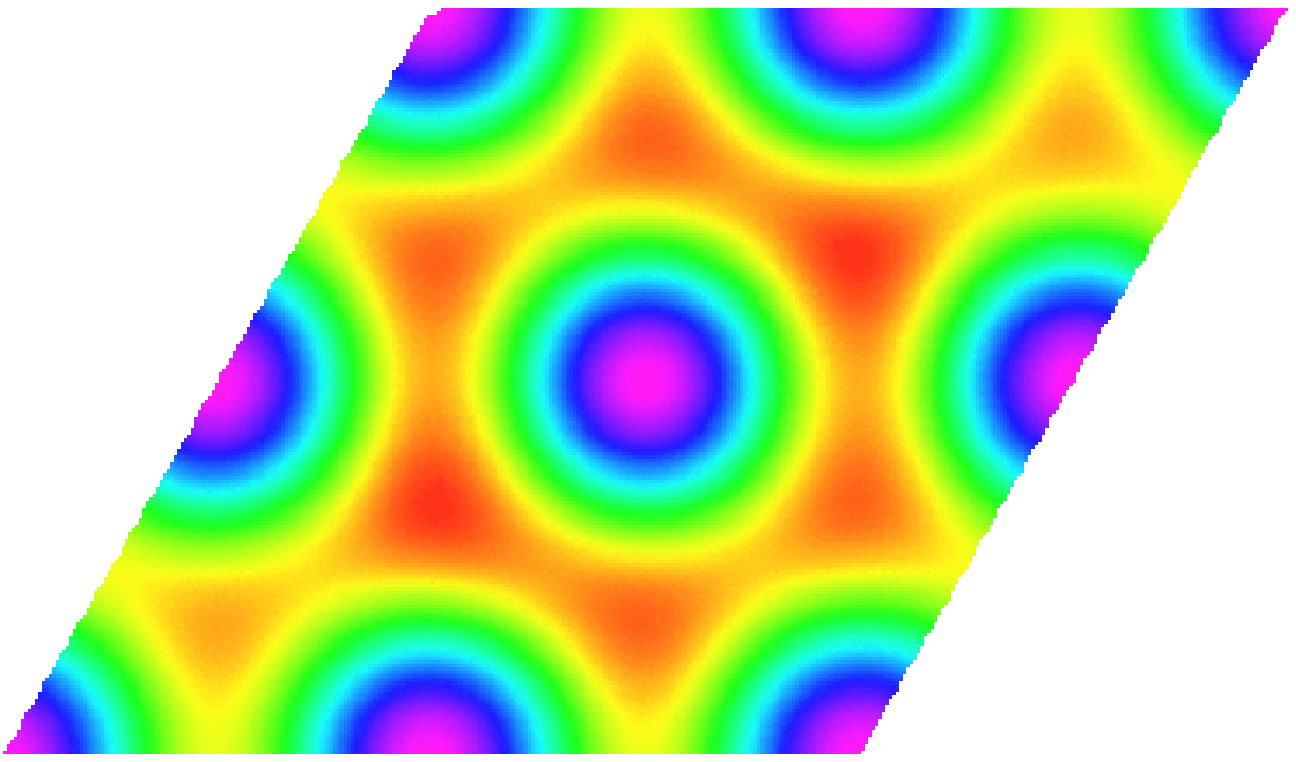}}
\\
\hline
\subfloat[$\rho=0.25$, $E/8 \pi =1.087$]{
\label{figureG2c}
\includegraphics[angle=0,scale=1,width=0.22\textwidth]{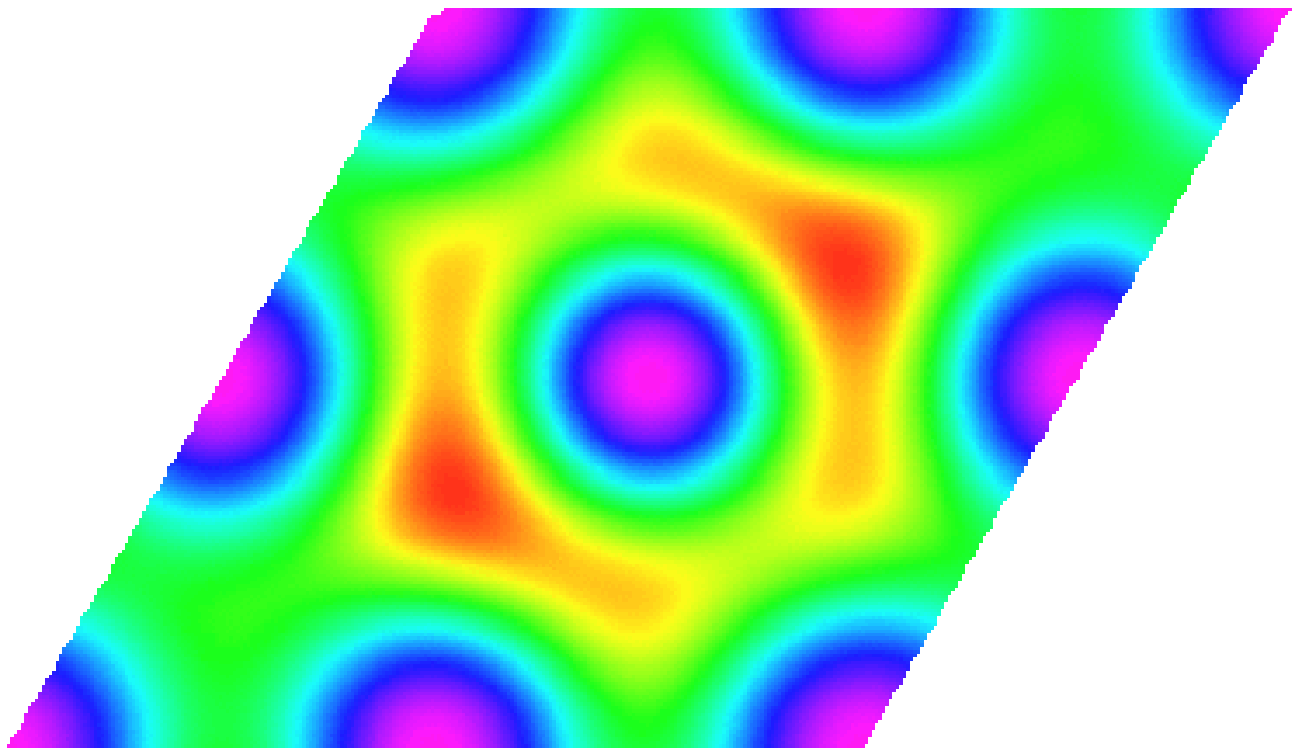}}
&
\subfloat[$\rho=0.14$, $E/8 \pi =1.08$]{
\label{figureG2d}
\includegraphics[angle=0,scale=1,width=0.22\textwidth]{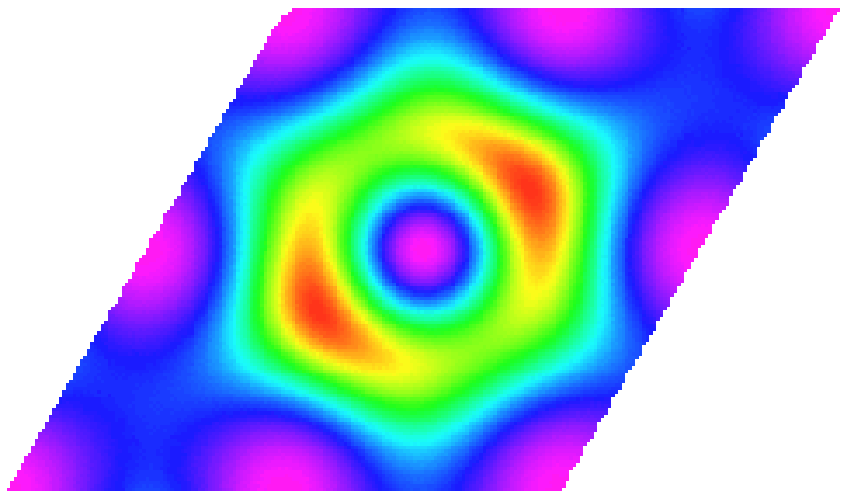}}
\\
\hline
\subfloat[$\rho=0.0625$, $E/8 \pi =1.082$]{
\label{figureG2e}
\includegraphics[angle=0,scale=1,width=0.22\textwidth]{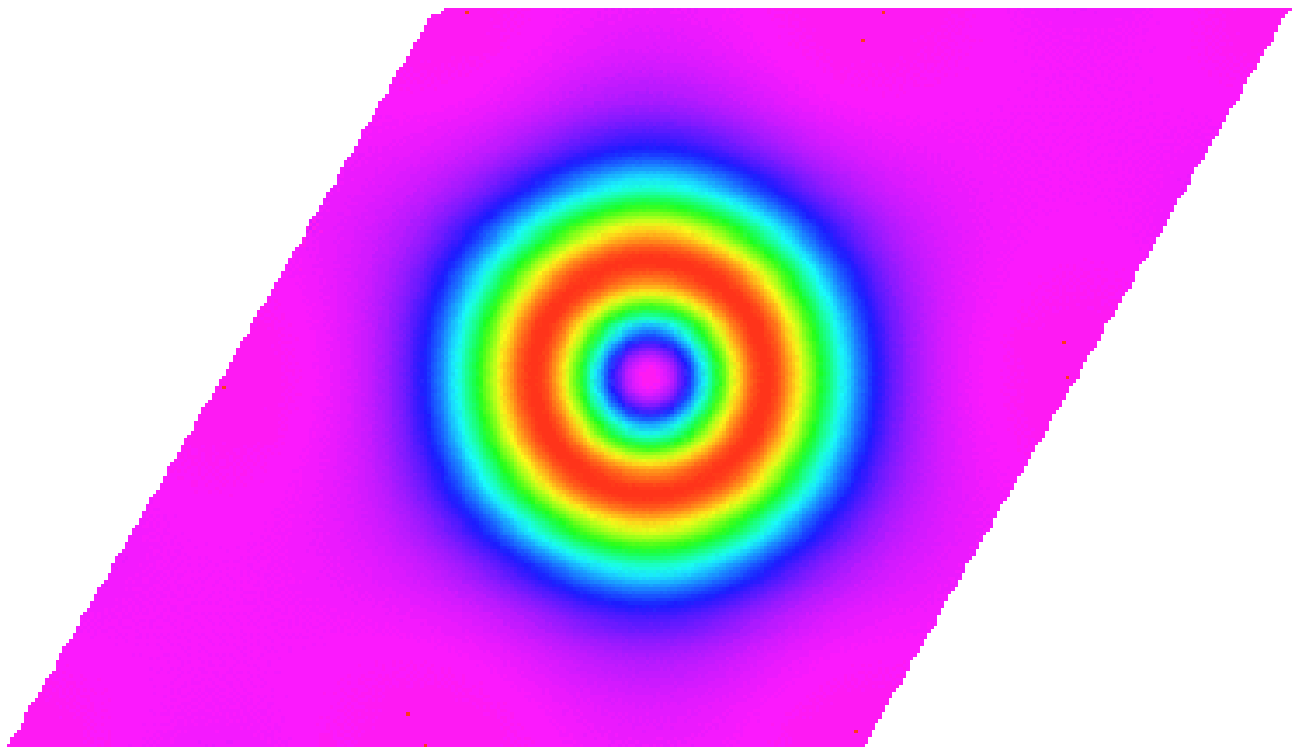}}
&
\subfloat[$\rho=0.03125$, $E/8 \pi =1.084$]{
\label{figureG2f}
\includegraphics[angle=0,scale=1,width=0.22\textwidth]{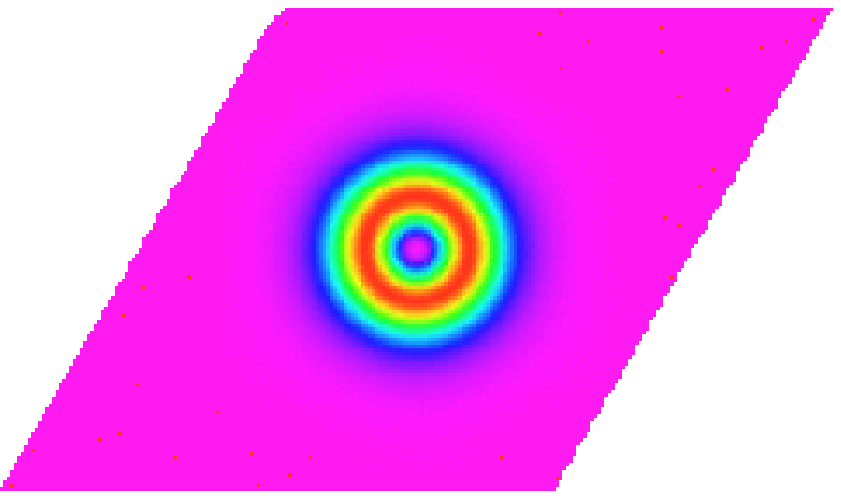}}
\\
\hline
\end{tabular}
\caption{\label{contourG2} Charge-two Skyrmions in the general case with \hbox{$\mu^2=0.1$} and $\kappa^2=0.03$: 
Contour plots of the charge densities of the minimal-energy configurations
in the hexagonal setting for different densities. Here,
the energetically most favorable density is $\rho= 0.14$.
The plot colors range from violet (low density) to red (high density).$\hfill{}$}
\end{figure}

Furthermore, due to the finite size of the Skyrmion, there is
an optimal density for which the energy is minimal among all densities. 
Figure \ref{contourG2} shows the contour plots of the charge density 
of the charge-two Skyrmion for several densities 
with $\kappa^2=0.03$ and $\mu^2=0.1$. The energy of the Skyrmion is minimal
for $\rho \approx 0.14$ (Fig. \ref{Energy}). 
\section{\label{sec:semiAna} Semi-analytical approach}
The energy functional (\ref{eq:T2energy}) depends both on the Skyrme field $\bphi$ and on the parallelogram parameters
$\gamma$ and $s$. Formally, the minimal energy configuration over all parallelograms may be obtained by 
functional differentiation with respect to $\bphi$ and regular
differentiation with respect to $\gamma$ and $s$.
However, since the resulting equations are very complicated, 
a direct numerical solution is quite hard.
Nonetheless, some analytical progress may be achieved in the following way. 
As a first step, we differentiate the energy functional (\ref{eq:T2energy}) only with respect to $\gamma$ and $s$:
\bea \label{eq:SGder}
\frac{\partial E}{\partial \gamma} &=& \frac{1}{2 s \cos^2 \gamma} \left(
\sin \gamma (\mathcal{E}^{yy} + s^2 \mathcal{E}^{xx}) -2 s \mathcal{E}^{xy} \right)  =0 \,, \nonumber\\*
\frac{\partial E}{\partial s}      &=&   \frac{1}{2 s^2 \cos \gamma} (\mathcal{E}^{yy} - s^2 \mathcal{E}^{xx}) =0 \,,
\eea
where $\mathcal{E}^{ij}=\int_{\mathbb{T}_2} \rmd x \rmd y  (\partial_{i} \bphi  \cdot \partial_{j} \bphi)$ and
$i,j \in \{ x,y \}$.
Solving these equations for $\gamma$ and $s$ yields
\bea \label{eq:sGammaMin}
s           &=& \sqrt{\frac{\mathcal{E}^{yy}}{\mathcal{E}^{xx}}} \,, \\\nonumber
\sin \gamma &=& \frac{\mathcal{E}^{xy}}{\sqrt{\mathcal{E}^{xx} \mathcal{E}^{yy}}} \,. 
\eea
Substituting these expressions 
into the energy functional (\ref{eq:T2energy}), we arrive at a `reduced' functional
\bea \label{eq:redE}
E= \sqrt{\mathcal{E}^{xx} \mathcal{E}^{yy} -{(\mathcal{E}^{xy})}^2} +\frac{\kappa^2 \rho}{2 B} \mathcal{E}_{\textrm{sk}} 
+\frac{\mu^2 B}{\rho} \mathcal{E}_{\textrm{pot}}\,,
\eea
where 
$\mathcal{E}_{\textrm{sk}}=\int_{\mathbb{T}_2} \rmd x \rmd y  \left( \partial_x \bphi \times \partial_y \bphi \right)^2$
is the Skyrme energy and $\mathcal{E}_{\textrm{pot}}=\int_{\mathbb{T}_2} \rmd x \rmd y  \left( 1-\phi_3 \right)$
is the potential energy.
Now that both $\gamma$ and $s$ are eliminated from the resultant expression, 
and the conditions for their optimization are built into the functional,
the numerical minimization is carried out.
We note here, however, that the procedure presented above should be treated with caution.
This is since the `minimization' conditions for $\gamma$  and $s$ in (\ref{eq:sGammaMin}) are
in fact only extremum conditions, and may turn out to be maximum or saddle-point conditions. 
Hence, it is important to confirm any results obtained using this method
by comparing them with corresponding results obtained from the method described in the previous section. 
\par
It is therefore reassuring that numerical minimization of the reduced functional (\ref{eq:redE})
gives $\sin \gamma = 0.498$ ($\gamma \approx \pi/6$) and $s=1$ (both for the Skyrme case and the general case), 
confirming the results presented in the previous section.
\par
In the general ($\mu \neq 0$) case, the energy functional (\ref{eq:redE}) may be further differentiated
with respect to the Skyrmion density
$\rho$ to obtain the optimal density for which the Skyrmion energy is minimal.
Differentiating with respect to $\rho$, and substituting the obtained 
expression into the energy functional, results in the functional 
\bea \label{eq:redE2}
E= \sqrt{\mathcal{E}^{xx} \mathcal{E}^{yy} -{(\mathcal{E}^{xy})}^2} 
+\kappa \mu \sqrt{ 2 \mathcal{E}_{\textrm{sk}} \mathcal{E}_{\textrm{pot}}}  \,.
\eea
Numerical minimization of the above expression for $\kappa^2=0.03$ 
and various $\mu$ values ($0.1 \leq \mu^2 \leq 10$)
yielded the hexagonal setting as in the Skyrme case.
In particular, for $\mu^2=0.1$ the optimal density 
turned out to be $\rho \approx 0.14$, in accord
with results presented in sub-section \ref{subsec:general}.

\section{\label{sec:summary} Summary and further remarks}
In this  paper, we have studied the crystalline 
structure of baby Skyrmions in two dimensions
by finding the minimal energy configurations
of baby Skyrmions placed inside parallelogramic fundamental unit cells
and imposing periodic boundary conditions.
In the pure $O(3)$ case (where both the Skyrme and potential terms are absent), we verified
that there are no favorable lattices in which the Skyrmions order themselves, as all
parallelogram settings yielded the same minimal energy,
saturating the minimal energy bound 
given by inequality (\ref{ineq}).
\par
In the Skyrme case, without the potential term,
the results are different.
For any fixed Skyrmion density,
the parallelogram for which the Skyrmion energy 
is minimal turns out to be the hexagonal lattice (for which $s=1$ and $\gamma = \pi/6$).
In particular, the hexagonal lattice  has lower energy than the four half-Skyrmions
configuration on the rectangular lattice.  
For example, at $\rho=2$, $E=1.433$ for the hexagonal lattice vs. $E=1.446$ 
for the rectangular lattice.
\par
The hexagonal setting turns out to be the energetically most favorable also in the general case
with both the Skyrme term and the potential term.
In this case, however, the model is not scale invariant and
the Skyrmion has a definite size. This results in 
the existence of a phase transition as a function of density and the appearance of an optimal density, 
for which the total energy of the Skyrmion is minimal.
This is analogous to the Skyrmion behavior in the $3$D Skyrme model,
in which the Skyrmions also possess a definite size \cite{Opt3d}.
\par
As pointed out in the Introduction,
the special role of the hexagonal lattice revealed here is not 
unique to the baby Skyrme model, but in fact arises 
in other solitonic models.
In the context of Skyrme models, the existence of a hexagonal two dimensional structure of $3$D Skyrmions
has been found by \cite{3Dhexlat}, where it has already been noted that 
energetically, the optimum infinite planar structure of $3$D Skyrmions 
is the hexagonal lattice, which resembles the 
structure of a graphite sheet, the most stable form of carbon thermodynamically
(\cite{TopoSol}, p. 384). 
Other examples in which the hexagonal structure is revealed are Ginzburg-Landau vortices which 
are known to have lower energy 
in a hexagonal configuration than in a square lattice configuration \cite{GL}.
Thus, it should not come as a surprise that the hexagonal structure 
is found to be the most favorable in the baby Skyrme model. 
\par
As briefly noted in the Introduction, a certain type of baby Skyrmions also arise in quantum Hall systems
as low-energy excitations of the ground state near ferromagnetic filling factors (notably $1$ and $1/3$) 
\cite{QHE}. It has been pointed out that this state contains a finite density of Skyrmions \cite{Brey},
and in fact the hexagonal configuration has been suggested as a candidate for their lattice structure \cite{QHF1}. 
Our results may therefore serve as a supporting evidence in that direction, 
although a more detailed analysis is in order.
\par
Our results also raise some interesting questions.
The first is how the dynamical properties of baby Skyrmions on the
hexagonal lattice differ from their behavior in the usual rectangular lattice.
Another question has to do with their behavior in non-zero-temperature settings.
\par 
One may also wonder whether and how these results can be generalized to 
 the $3$D case, once  a systematic study like the one reported here is conducted.
Is the face-centered cubic lattice indeed the minimal energy crystalline structure of $3$D Skyrmions
among all parallelepiped lattices? If not, what would the minimal energy structure be? 
and how would these results change when a mass term is present? 
We hope to answer these questions in future research. 

\begin{acknowledgments}
We thank Alfred S. Goldhaber, Moshe Kugler, Philip Rosenau and Wojtek Zakrzewski for useful discussions
and suggestions. 
This work was supported in part by a grant from the Israel Science
Foundation administered by the Israel Academy of Sciences and Humanities.
\end{acknowledgments}

\end{document}